\documentstyle[11pt]{article}
\input amssymb.sty
\newcommand{\tr}{{\rm tr}}
\newcommand{\ad}{{\rm ad}}

\newcommand{\ti}[1]{\tilde{#1}}
\newcommand{\om}{\omega}
\newcommand{\Om}{\Omega}
\newcommand{\de}{\delta}
\newcommand{\al}{\alpha}
\newcommand{\te}{\theta}
\newcommand{\vth}{\vartheta}

\newcommand{\La}{\Lambda}

\newcommand{\ve}{\varepsilon}

\newcommand{\vf}{\varphi}
\newcommand{\G}{\Gamma}

\newcommand{\ga}{\gamma}
\newcommand{\ze}{\zeta}
\newcommand{\si}{\sigma}

\def\bfc{{\bf c}}

\def\bfe{{\bf e}}

\def\bfu{{\bf u}}
\def\bfv{{\bf v}}

\def\bfS{{\bf S}}
\def\bfJ{{\bf J}}
\def\bv{{\bf v}}

\def\cS{{\cal S}}
\def\cA{{\cal A}}
\def\mC{{\mathbb C}}
\def\mZ{{\mathbb Z}}
\def\mR{{\mathbb R}}
\def\mN{{\mathbb N}}
\def\rot{{\rm curl}}
\newcommand{\ar}[2]{\!\left[\!\begin{array}{c} {#1}\\{#2}\end{array}\!\right]}

\newcommand{\beq}[1]{\begin{equation}\label{#1}}
\newcommand{\eq}{\end{equation}}
\newcommand{\beqn}[1]{\begin{eqnarray}\label{#1}}
\newcommand{\eqn}{\end{eqnarray}}
\newcommand{\p}{\partial}
\newcommand{\di}{{\rm diag}}
\newcommand{\oh}{\frac{1}{2}}

\newcommand{\slt}{{\rm sl}(2,{\mathbb C})}
\newcommand{\GLN}{{\rm GL}(N,{\mathbb C})}
\def\sln{{\rm sl}(N, {\mathbb C})}
\def\SLN{{\rm SL}(N, {\mathbb C})}
\newcommand{\gln}{{\rm gl}(N, {\mathbb C})}

\def\f1#1{\frac{1}{#1}}

\newcommand{\bp}{\bar{\partial}}
\newcommand{\bz}{\bar{z}}

\def\frak{\mathfrak}
\def\gg{{\frak g}}
\def\gJ{{\frak J}}
\def\gS{{\frak S}}

\def\cS{{\cal S}}
\def\cA{{\cal A}}

\newcommand{\ran}{\rangle}
\newcommand{\lan}{\langle}

\renewcommand{\theequation}{\thesection.\arabic{equation}}

\newcommand{\AmS}{{\protect\the\textfont2
  A\kern-.1667em\lower.5ex\hbox{M}\kern-.125emS}}

\begin{document}
% declarations for front matter

\vspace{0.3in}
\begin{flushright}
 ITEP-TH-29/03\\
%{\large{\it Very preliminary draft}}\\
\today
\end{flushright}
\vspace{10mm}
\begin{center}
{\Large{\bf
Large $N$ limit of integrable models}
}\\
\vspace{5mm}
M. Olshanetsky
\\
{\sf Institute of Theoretical and Experimental Physics, Moscow, Russia,}\\
{\em e-mail olshanet@gate.itep.ru}\\

\vspace{5mm}
\end{center}
\begin{flushright}
{\sl Dedicated to the sixty-fifth birthday of Sergey Novikov}
\end{flushright}
\vspace{10mm}

\begin{abstract}
We consider a large $N$ limit of the
Hitchin type integrable systems. The first system is the elliptic rotator on $GL_N$
that corresponds to the Higgs bundle of degree one over an
elliptic curve with a marked point. This system is gauge equivalent
to the $N$-body elliptic Calogero-Moser system, that is derived
from the Higgs bundle of degree zero over the same curve. The
large $N$ limit of the former system is the integrable rotator on
the group of the non-commutative torus. Its classical limit
 leads to the integrable modification of 2d hydrodynamics
on the two-dimensional torus. We also consider the elliptic Calogero-Moser
system on the group of the non-commutative torus and consider
the systems that arise after the reduction to the loop group.
\vspace{1pc}
\end{abstract}

\vspace{0.15in}
\bigskip
\title
\maketitle
\tableofcontents

\section{Introduction}
\setcounter{equation}{0}

In this paper we analyze two related integrable models - a modified integrable
two-dimensional hydrodynamics on a torus and the large $N$ limit
of the elliptic Calogero-Moser system (ECMS) with spin.
The former system is also the large $N$ limit of the integrable
$\GLN$ elliptic rotator (ER) proposed in Ref.\,\cite{STSR}.
Integrable models are among the main scientific interests of Sergey Novikov
and his contribution to this subject is widely recognized.

It was established in Ref.\,\cite{LOZ} that for finite $N$ ECMS
is gauged equivalent to ER. The
both systems are particular examples of the Hitchin construction \cite{Hi,Mar,Ne}. Namely,
they are derived from the Higgs bundles of rank $N$
over an elliptic curve. The corresponding bundle for ECMS has
degree zero while the bundle corresponding to ER has degree
one. The gauge equivalence is determined by the upper modification that
transforms the trivial bundle into the bundle of degree one.

We analyze the both systems in a similar way trying to establish
 the structures that were already
known for the open finite-dimensional Toda chain. Namely, for
the open finite-dimensional Toda chain there exists a limit to the infinite
chain \cite{UT}, reduction of the infinite chain to the periodic, and the dispersionless version
of the infinite chain \cite{SV,TT}.

 Here we consider a special limit  $N\to\infty$  that corresponds
to the passage from $\GLN$ to the infinite-dimensional group of the non-commutative torus (NCT).
We consider first ER
\footnote{This part is an extended version of the talk\, \cite{O}.}. This system is an example of
the integrable Euler-Arnold tops (EAT)
on the group $\SLN$. EAT are Hamiltonian systems defined
on coadjoint orbits of groups \cite{Ar}. Particular examples of
such systems are the Euler top on SO(3), its integrable SO($N$) generalization
\cite{Di,Man} and  the hydrodynamics  of the ideal incompressible
fluid on a space $M$.
The corresponding group of the latter system is SDiff$(M)$.
We consider here the case when $M$ is a torus $T^2$.
EAT are completely determined by their Hamiltonians,
since the Poisson structure is fixed to be related to
 the Kirillov-Kostant form on the coadjoint orbits.
The Hamiltonians are determined by the inertia-tensor operator $\gJ$
mapping the Lie algebra $\gg$ to the Lie coalgebra $\gg^*$.
 Special choices of $\gJ$ lead to completely integrable systems
(see review \cite{DKN}). In the case of
the 2d hydrodynamics $\gJ$ has the form of the Laplace operator and it
turns out that the theory is non-integrable \cite{Z}. One of the goals
of this paper are integrable EAT related to \mbox{SDiff}.
Some integrable models related to SDiff were considered in
\cite{TT,GKR,T}.

Integrable models on  SDiff$(M)$  can be described as the classical limit
(dispersionless limit) of integrable models when the commutators in the Lax equations
are replaced by Poisson brackets. This approach was
proposed in Ref.\,\cite{LM,Za}, and later developed in numerous
publications (see, for example, the review \cite{TT1}).

Here we use the same strategy defining an integrable system on
the non-commutative torus (NCT) and then taking the classical
limit to SDiff$(T^2)$.
In analogy with ER on $\GLN$
we consider a special limit $N\to\infty$ of $\GLN$ that leads to
 the group $G_\te$ of the NCT, where $\te$ is the non-commutative
parameter.
\footnote{For Manakov's top $\lim N\to\infty$
was considered in Ref.\cite{War}.}
The Hamiltonian is defined by
the  inertia-tensor operators depending on the module
$\tau,~Im\tau>0$ of an elliptic curve. This curve is the basic
spectral curve in the Hitchin description of the model.
The group $G_\te$ is defined as the
set of invertible elements of the  NCT algebra ${\cal A}_\te$.
It can be embedded in GL$(\infty)$ and in this way $G_\te$
can be described as a special limit of $\GLN$. We define a family of
EAT on $G_\te$ parameterized by $\tau$.
Then, we construct the Lax operator with the spectral parameter
on an elliptic curve with the same parameter $\tau$.

In the classical limit $\te\to 0$, $G_\te\to$SDiff$(T^2)$ and the
inertia-tensor operator $\gJ$ takes the form $\bp^2$. The conservation
laws survive in this limit while commutators in the Lax hierarchy
become the Poisson brackets. It turns out that
 the classical limit is essentially the same
as the rational limit of the basic elliptic curve, so that the product
of the Planck constant $\te$ and the half periods of the basic curve
are constant.

We also construct ECMS system related to NCT. In both cases  of we discuss  the
systems that arise after the reduction to the loop algebra
$\hat{L}(\GLN)$.

\section{The Lie algebra of the non-commutative torus}
\setcounter{equation}{0}

Here we reproduce some basic results about NCT and
related to it the Lie algebra $sin_\te$ \cite{FFZ}.

{\sl 1. Non-commutative torus.}

NCT ${\cal A}_\te$ is an unital algebra with
 two generators $(U_1,U_2)$ that satisfy the relation
\beq{3.1}
U_1U_2=\bfe(-\te) U_2U_1,~\bfe(\te)=e^{ 2\pi i \te},~ \te\in[0,1)\,.
\eq

Elements of ${\cal A}_\te$ are the double sums
$$
{\cal A}_\te=
\{x=\sum_{m,n\in{\mathbb Z}}a_{m,n}U_1^mU_2^n,~a_{m,n}\in\mathbb C\}\,,
$$
where $a_{m,n}$ are elements of the ring ${\gS}$
of the rapidly decreasing sequences on ${\mathbb Z}^2$
\beq{3.1a}
\gS=\{a_{m,n}\,|\,
{\rm sup}_{m,n\in\mZ}(1+m^2+n^2)^k|a_{m,n}|^2<\infty,\,
{\rm for~all}~k\in \mN\}\,.
\eq
The trace functional $\tr(x)$ on ${\cal A}_\te$ is defined as
\beq{tr}
\tr(x)=a_{00}\,.
\eq
The dual space to $\gS$ is
\beq{ds}
\gS'=\{s_{k,j}~|~\sum_{m+j=0,~n+k=0} a_{m,n}s_{kj}<\infty,~~a_{m,n}\in\gS\}\,.
\eq

The associative algebra ${\cal A}_\te$ can be regarded as the quantization
of the commutative algebra of smooth functions
on the two-dimensional torus
\beq{T}
T^2=\{\mR^2/\mZ\oplus\mZ\}\,\sim \,\{0<x\leq 1,\,0<y\leq 1\}.
\eq
by means of the identification
\beq{3.2}
U_1\to\bfe(x),~~U_2\to\bfe(y),
\eq
where the multiplication of functions on $T^2$ is the Moyal multiplication:
\beq{3.3}
(f*g)(x,y):=fg+
\sum_{n=1}^\infty\frac{(\pi\te)^n}{n!}
\ve_{r_1,s_1}\ldots\ve_{r_n,s_n}(\p^n_{r_1\ldots r_n}f)
(\p^n_{s_1\ldots s_n}g).
\eq
The trace functional (\ref{tr}) in the Moyal identification
is the integral
\beq{3.6}
\tr f=\int_{{\cal A}_\te}fdxdy=f_{00}\,.
\eq

Another representation of ${\cal A}_\te$ is defined by the operators,
that act on the space of Schwartz functions on $\mR$
\beq{3.4}
U_1\to\bfe(-2\pi\te\p_\vf),~U_2\to\exp(i\vf)\,.
\eq

Finally, we can identify $U_1,U_2$ with matrices from GL$(\infty)$.
We define GL$(\infty)$ as the associative algebra of infinite matrices
$a_{jk}E_{jk}$ such
that
$$
{\rm sup}_{j,k\in\mZ}|a_{jk}|^2|j-k|^n<\infty\, ~{\rm for~all}~n\in\mN\,.
$$
Consider the following two matrices from GL$(\infty)$
$$
Q=\di (\bfe(j\te)),~\La=E_{j,j+1},~j\in\mZ\,.
$$
We have the following identification
\beq{3.5}
U_1\to Q,~U_2\to \La\,.
\eq

\bigskip

{\sl 2. sin-algebra}

Define the following quadratic combinations of the generators
\beq{3.10}
T_{m,n}=\frac{i}{2\pi\te}\bfe
\left(\frac{mn}{2}\te
\right)U_1^mU_2^n\,.
\eq
Their commutator has the form of the sin-algebra \cite{FFZ}
\beq{3.11}
[T_{m,n},T_{m'n'}]
=\f1{\pi\te}\sin\pi\te(mn'-m'n)T_{m+m',n+n'}\,.
\eq
We denote by $sin_\te$ the Lie algebra with the generators $T_{m,n}$
over the ring $\gS$ (\ref{3.1a})
\beq{3.12}
\psi=\sum_{m,n}\psi_{m,n}T_{m,n},~~\psi_{m,n}\in\gS\,,
\eq
 and by $G_\te$ the group of invertible elements
from ${\cal A}_\te$. The coalgebra $sin^*_\te$ is the linear space
$$
sin^*_\te=\{{\mathcal S}=\sum_{jk}s_{jk}T{jk},~s_{jk}\in\gS'\}\,.
$$

In the Moyal representation (\ref{3.3}) the commutator of $sin_\te$
has the form
\beq{3.14}
[f(x,y),g(x,y)]=\{f,g\}^*:=\f1{\te}(f*g-g*f)\,.
\eq

The algebra $sin_\te$ has a central extension $\widehat{sin}_\te$.
 The corresponding
additional term in (\ref{3.11}) has the form of the star-brackets
\beq{3.13}
(am+bn)\de_{m,-m'}\de_{n,-n'}\,,~~~a,b\in\mC\,.
\eq
In other words, the commutator in $\widehat{sin}_\te$ takes the form
$$
[f,g]=\{f,g\}^*+\bfc\f1{4\pi^2}\int_{{\cal A}_\te}f(a\p_xg+b\p_yg)\,.
$$

\bigskip
{\sl 3. Loop algebra} $\hat{L}(\sln)$.

Let $\te$ be a rational number $\te=p/N$, where $p,N\in\mN$
are mutually prime. In this case
${\cal A}_\te$ has the ideal
$$
I_N=\{\sum c_{m,n}^{(l)}(U_1^mU_2^n-U_1^{m+Nl}U_2^n)=0\,,~~l=\overline{1,N}\}\,.
$$
The factor-algebra ${\cal A}_\te/I_N$
can be represented by embedding in GL$(\infty)$. Represent an arbitrary
element of GL$_\infty$ as
$$
\psi_{m.n}\bfe(\frac{mn}{N})U_1^mU_2^n\,.
$$
In the factor-algebra one has $\psi_{Ns+k,n}=\psi_{k,n}$. Then any element
from ${\cal A}_\te/I_N$ takes the form
$$
\sum_{l\in\mZ}a_{j,r}^{(l)}E_{j,j+Nl+r}\,,~~j=\overline{1,N}\,,
r=\overline{-N+1,N-1}\,,
$$
where $a_{j,r}^{(l)}=\sum_{k=1}^N\psi_{k,Nl+r}\bfe(\frac{kj}{N})$.
We put in correspondence the current from $L(\sln)$
$$
g(t)=\sum_{l\in\mZ}g_{j,r}^{(l)}E_{j,j+r}t^{Nl+r}\,.
$$
 After the gauge transform by
$\di(1,t,\ldots,t^{N-1})$ we kill the factor $t^r$ and then by replacing $w=t^N$ we
come to the loop algebra with the principle gradation
$$
g(w)=\sum_{l\in\mZ}g^{(l)}w^l\,.
$$
The central extension $\oint Tr(g_1(w)\p_wg_2(w)\frac{dw}{w}$
is proportional to the cocycle (\ref{3.13}) for $a=0$, $b=1$. Here $Tr$ is the trace in
the fundamental representation of $\sln$.

\section{2d-hydrodynamics on ${\cal A}_\te$}
\setcounter{equation}{0}

{\sl 1. 2-d hydrodynamics}

Let $\bv=(V_x,V_y)$ be the velocity of the ideal incompressible
fluid on a compact manifold $M$ (dim$(M)=2$, div$\bv=0$) and
$\rot \bv=\p_xV_y-\p_yV_x$ be its
 vorticity.
\footnote{For simplicity we assume that the measure on $M$ is $dx\wedge dy$,
though all expressions can be written in a covariant way.}
The Euler equation for 2d hydrodynamics   takes
the form \cite{Ar}
\beq{Ar1}
\p_t\rot \bv=\rot[\bv,\rot \bv]\,.
\eq
Define the stream function $\psi(x,y)$ as
 the Hamiltonian function generating the vector field $\bv$
$$
i_\bv dx\wedge dy=d\psi\,.
$$
In other words
\beq{VH}
V_x=\p_y\psi\,,~~V_y=-\p_x\psi\,.
\eq

Let $\gg$ be the Poisson
algebra of the stream functions $\gg=\{\psi\}$ on $M$
defined up to constants $\gg\sim C^\infty(M)/\mC$
$$
\{\psi_{\bv_1},\psi_{\bv_2}\}=-i_{\bv_1}d\psi_2.=\,.
$$
Consider the Lie algebra $SVect{M}$ of vector fields div$\bv=0$
 on $M$. We have the following interrelation between the Lie
algebras $\gg$ and $SVect(M)$
$$
\psi_{[\bv_1,\bv_2]}=\{\psi_{\bv_1},\psi_{\bv_2}\}\,,
$$
$$
0\to\mC^2\to SVect(M)\to\gg\to 0\,,
$$
where the map $ SVect(M)\to\gg$ is defined by (\ref{VH})
and the image of $\mC^2$ is generated by the two fluxes
$(c_1\p_1,c_2\p_2)$.

 Let $\gg^*$ be the dual space of
distributions on $M$.
The vorticity $\cS=\rot \bv$
of the vector field $\bv$
 $$
\cS=-\Delta\psi
$$
can be considered as an element from $\gg^*$.
 The Euler equation (\ref{Ar1}) in terms of the Poisson brackets  has the form
\beq{Ar2}
\p_t\cS=\{\cS,\psi\},~{\rm or}~~\p_t\cS=\{\cS,\Delta^{-1}\cS\}\,.
\eq

We can view (\ref{Ar2}) as the Euler-Arnold equation
for the rigid top related
to the Lie algebra $\gg$, where the Laplace operator is the map
$$
\Delta \,: \gg\to\gg^*
$$
that plays the role of the inertia-tensor. The phase space of the system
is a coadjoint orbit of the group of the canonical transformations
 SDiff$(M)$.
The equation (\ref{Ar2}) takes the form
\beq{Ar3}
\p_t{\cal S}=\ad^*_{\nabla H}{\cal S}\,,
\eq
where ${\nabla H}=\frac{\de H}{\de \cS}=\psi$
is the variation of the Hamiltonian
\beq{Ar4}
H=-\oh\int_M\cS\Delta^{-1}\cS=\int_M\psi\Delta\psi\,.
\eq
There is infinite set of Casimirs defining the coadjoint
orbits:
\beq{Ca}
C_k=\int_M\cS^k\,.
\eq

Consider a particular case, when $M$ is a two-dimensional torus (\ref{T})
equipped with the measure $-\frac{dxdy}{4\pi^2}$.
In terms of the Fourier modes $s_{m,n}$ of the vorticity
$$
\cS=\sum_{m,n}s_{m,n}\bfe(-mx-ny)
$$
the Hamiltonian (\ref{Ar4}) is
\beq{Ar6}
H=-\oh\sum_{m,n}\f1{m^2+n^2}s_{m,n}s_{-m,-n}\,,
\eq
and we come to the equation
\beq{Ar5}
\p_ts_{m,n}=\sum_{j,k}\frac
{jn-km}{j^2+k^2}s_{jk}s_{m-j,n-k}\,.
\eq

{\sl 2. 2d hydrodynamics  on non-commutative torus}.

We can consider the similar construction by replacing the
Poisson brackets by the Moyal brackets (\ref{3.14}) \cite{Ze,HOT}.
Introduce the vorticity $\cS$ as an element of $sin^*_\te$
\beq{4.00}
\cS=\sum_{m,n}s_{m,n}T_{-m,-n}.
\eq
The equation (\ref{Ar2}) takes the form
$$
\p_t\cS(x,y)= \{\cS(x,y),\Delta^{-1}\cS(x,y)\}^*\,,
$$
or for the Fourier modes
\beq{Ar7}
\p_ts_{m,n}=
\f1{8\pi^3\te}\sum_{j,k}\frac
{\sin(\pi\te(jn-km))}{j^2+k^2}s_{j,k}s_{m-j,n-k\,}\,.
\eq
This system is EAT on the group $G_\te$ of
invertible elements of ${\cal A}_\te$ and the coadjoint orbits
are defined by the same Casimirs (\ref{Ca}) as for SDiff$(T^2)$.
In the limit $\te\to 0$ (\ref{Ar7}) reproduces (\ref{Ar5}).

\section{$\SLN$-elliptic rotator}
\setcounter{equation}{0}

{\sl 1. Elliptic rotator (ER) on $\SLN$}

Now we consider differential equations related to $\SLN$ apriori
not coming from the hydrodynamics.
The elliptic  $\SLN$-rotator  is an example of EAT \cite{Ar}.
It is defined on $\sln^*$ and its phase space is
 a coadjoint orbit of $\SLN$:
\beq{8.4}
{\mathcal R}^{rot}=\{\cS\in\sln,~~\cS=g^{-1}\cS^{(0)}g\}\,.
\eq
The phase space ${\mathcal R}^{rot}$ is equipped with the
Kirillov-Kostant symplectic form
\beq{fo}
\om^{rot}=\tr(\cS^{(0)}Dgg^{-1}\wedge Dgg^{-1})\,.
\eq
 The Hamiltonian has the form
\beq{8.5}
H^{rot}=-\oh Tr(\cS J(\cS))\,,
\eq
where $J$ is a linear operator on  $\sln$. The inverse
operator is called the inertia tensor.
The equation of motion takes the form
\beq{em}
\p_t\cS=[\cS,J(\cS)].
\eq

A special form of $J$ provides the integrability
of the system \cite{STSR,LOZ}. Represent $\cS$ in the form
 $$
\cS =-\frac{4\pi^2}{N^3}\sum_{m,n}S_{m,n}T_{-m,-n}\,,
$$
 where
 $T_{m,n}$ is the basis of $\sln$, similar to
the basis of the sin-algebra (\ref{3.10})
$$
T_{m,n}=\frac{iN}{2\pi}\bfe(\frac{mn}{N})Q_N^m\La_N^n\,,
$$
$$
(m,n)\in\ti\mZ^{(2)}_N=\{(\mZ/N\mZ\oplus\mZ/N\mZ)\setminus (0,0)\}\,.
$$
Here $\La_N\,,Q_N$ are defined by (\ref{ql}).
Let
$$
J(\cS)=\sum_{m,n\in\mZ} J_{m,n}S_{m,n}T_{-m,-n}\,,
$$
 where
\beq{8.6}
\bfJ=\{J_{m,n}\}=
\left\{\wp\left[\begin{array}{c}m\\n\end{array}\right]\right\},~~
~~\wp\left[\begin{array}{c}m\\n\end{array}\right]=
\wp\left(\frac{m+n\tau}{N};\tau\right)\,.
\eq
Then (\ref{em}) takes the form
\beq{8.5a}
\p_tS_{m,n}=\sum_{k,l\in\mZ}S_{k,l}S_{m-k,n-l}
\wp\left[\begin{array}{c}k\\l\end{array}\right]
\sin\left(\frac{\pi}{N}(ml-kn)\right)\,.
\eq

It was observed in Ref.\,\cite{Mar} that
ER is a Hitchin system corresponding to the vector
bundle $E$ of rank $N$ and degree one over the elliptic curve $E_\tau$
with the marked point $z=0$. To prove this fact we first demonstrate
that (\ref{8.5a}) is equivalent to the Lax equation
$$
\p_tL^{rot}(z)=[L^{rot}(z),M^{rot}(z)]\,,
$$
and then show that $ L^{rot}(z)$ is the Higgs field in corresponding bundle
(see Appendix A).
The Lax matrices in the basis $T_{m,n}$ of $\gln$
are represented as
\beq{8.7}
L^{rot}=\sum\limits_{m,n\in\mZ }
S_{m,n}\varphi\left[\begin{array}{c}m\\n\end{array}\right]\!\!(z)
T_{m,n},~~~
\varphi\left[\begin{array}{c}m\\n\end{array}\right]\!\!(z)=
\bfe(\frac{nz}{N})\phi(\frac{m+n\tau}{N};z)\,,
\eq
\beq{8.8}
M^{rot}=-\sum\limits_{m,n\in\mZ }
S_{m,n}f\left[\begin{array}{c}m\\n\end{array}\right]\!\!(z)T_{m,n},~~
f\left[\begin{array}{c}m\\n\end{array}\right]\!\!(z)=
\bfe(\frac{nz}{N})\partial_{u}\phi(u;z)|_{u=\frac{m+n\tau}{N}}\,.
\eq
They lead to the Lax equation for the matrix elements
$$
\p_tS_{m,n}\varphi\left[\begin{array}{c}m\\n\end{array}\right](z)=
\sum_{k,l\in\mZ }
S_{m-k,n-l}S_{kl}\varphi\left[\begin{array}{c}m-k\\n-l\end{array}\right]\!\!(z)
f\left[\begin{array}{c}k\\l\end{array}\right]\!\!(z)
\sin\frac{\pi}{N}(nk-ml)\,.
$$
Using the Calogero functional equation (\ref{ad2})
we rewrite it in the form (\ref{8.5a}).

The phase space ${\mathcal R}^{rot}$ (\ref{8.4})
is the result of the Hamiltonian reduction of the GL$_N$ Higgs bundle of degree one.
In this case there is no moduli degrees of freedom  except the Jacobian of the
determinant bundle (\ref{B4}).
In fact, the determinant bundle coincides with the theta-bundle and therefore
has degree one.
It implies that the gauge fixing is complete and
the reduced phase space is the orbit ${\cal O}$ (see (\ref{b6})).
 In the symplectic form (\ref{B7})
survives only the last term that coincides with (\ref{fo}).
For a generic orbit $\dim{\cal R}^{(1)}=N(N-1)$.
The transition
functions can be chosen in the form
(\ref{B15}).
The transition functions
(\ref{B15}) allow us to define the Lax operator depending only on
the orbit variables.
It can be checked directly that the
Lax operator (\ref{8.7}) is a meromorphic $\gln$-valued one form on $E_\tau$
$$
{\rm Res}L^{rot}|_{z=0}=\cS\,,
$$
that satisfies the quasi-periodicity conditions with the transition
functions (\ref{ql})
$$
L^{rot}(z)- Q_N L^{rot}(z+1) Q_N^{-1}=0\,,~~
L^{rot}(z)- \La_N L^{rot}(z+\tau) \La_N ^{-1}=0\,.
$$

It follows from the general prescription that
we have $\frac{N(N-1)}{2}$ independent integrals of motion (\ref{B12a}).
In particular,
$$
\frac{1}{2}\tr(L^{rot}(z))^2=-\frac{i\pi}{N}H^{rot}+\tr\cS^2\wp(z)\,.
$$

 The equations
of motion, corresponding to the higher integrals has the Lax form
(\ref{lax}).
The properties of $M_{s,j}(z)$ can be read of from the equation of
motion (\ref{B13a}) restricted to  ${\cal R}^{red}$
$$
M_{s,j}(z)- Q_NM_{s,j}(z+1) Q_N^{-1}=0\,.
$$
For $s=0$ we have
\beq{Ma}
M_{0,j}(z)- \La_N M_{0,j}(z+\tau) \La_N ^{-1}=
2\pi i(L)^{j-1}(z)\,.
\eq
If $s\neq 0$ then $M_{s,j}(z)$ is quasi-periodic
$$
M_{s,j}(z)- \La_N M_{s,j}(z+\tau) \La_N ^{-1}=0\,,
$$
and its singular part is defined by the singular
part of $L_N^{j-1}z^s$
\beq{Ma1}
\left(M_{s,j}(z)\right)_-=\left(L_N^{j-1}z^s\right)_-\,.
\eq

\section{Elliptic rotator on ${\cal A}_\te$}
\setcounter{equation}{0}

{\sl 1. Description of the system.}

It follows from (\ref{3.5}) that the non-commutative torus ${\cal A}_\te$
corresponds to a special limit $N\to\infty$ of the $\SLN$. Consider ER related
to the group of NCT $G_\te$  and assume that $\te$ is a irrational number.
We replace the inverse inertia-tensor $\Delta^{-1}$ of the hydrodynamics
on the operator $\bfJ:~\cS\to\psi$
acting in a diagonal way on the Fourier coefficients (\ref{4.00}):
\beq{4.1}
\bfJ:~s_{m,n}\to
\cdot s_{m,n}=\psi_{m,n}\,, ~~(s_{00}=0)\,,~~
 \wp\!\left[\!\begin{array}{c} m\\n\end{array}\!\right]=
\wp\left((m+n\tau)\te;\tau\right)\,.
\eq

We consider EAT on the group
$ G_\te$ with the inertia-tensor defined by
$\bfJ^{-1}$ (\ref{4.1}).
The corresponding coadjoint orbit is
\beq{co}
{\cal O}_{\cS^0}=\{\cS\in sin_\te^*~|~\cS=h^{-1}\cS_0h,~h\in G_\te,
~ \cS_0\in sin_\te^*\}
\eq
equipped with the Kirillov-Kostant symplectic form
$$
\om_\te=\int_{{\cal A}_\te}{\cS^0}Dhh^{-1}\wedge Dhh^{-1}\,.
$$
The Poisson structure on the coalgebra $sin_\te^*$ is defined by the Moyal
brackets
$$
\{\cS,\cS'\}=\{\cS,\cS'\}^*\,.
$$

Let
$$
\cS=-4\pi^2\te^3\sum_{m,n\in\mZ}s_{m,n}T_{-m,-n}\in
sin^*_\te\,,~~(s_{0,0}=0)\,,
$$
and
$$
{\bf J}(\cS)=\sum_{m,n\in\mZ}s_{-m,-n}
\wp\!\left[\!\begin{array}{c} m\\n\end{array}\!\right]
T_{m,n}\in sin_\te\,.
$$
 The Hamiltonian is determined by the integral over ${\cal A}_\te$ (\ref{3.6})
\beq{4.3}
H_\te=-\oh\int_{{\cal A}_\te}\cS \bfJ(\cS)=
-\frac{1}{2}\sum_{m,n\in\mZ }
 \wp\!\left[\!\begin{array}{c} m\\n\end{array}\!\right]s_{m,n}s_{-m,-n}\,.
\eq
We define the phase space below assuming now that the Hamiltonian $H_\te$ is
finite.
The equation of motion has the
standard form of the Moyal brackets, or the commutator in GL$(\infty)$
\beq{4.4}
\p_t\cS=\{\cS,\bfJ(\cS)\}^*=[\cS,\bfJ(\cS)]\,.
\eq
In the Fourier components it takes the form
\beq{4.2}
\p_ts_{m,n}=\f1{\pi\te}\sum_{j,k\in\mZ }s_{jk}s_{m-j,n-k}
\times\wp\!\left[\!\begin{array}{c} j\\k\end{array}\!\right]
\sin\left(\pi\te(jn-km)\right)\,.
\eq

\bigskip
{\sl 2. Integrability of elliptic rotator on ${\cal A}_\te$ }.

We will prove that the Hamiltonian system of ER
(\ref{4.4}), (\ref{4.2})  has an infinite set of involutive integrals of
motion in addition to the Casimirs (\ref{Ca}). It will follow from the Lax form
\beq{4.5}
\p_tL_\te=[L_\te,M_\te]
\eq
of the equations (\ref{4.4}), (\ref{4.2}).
The Lax operators are similar to the corresponding Lax matrices for the elliptic
rotator (\ref{8.7}), (\ref{8.8})
\beq{4.9}
L_\te=\sum_{mn\in\mZ }
s_{m,n}\vf\!\left[\!\begin{array}{c} m\\n\end{array}\!\right]\!\!(z)T_{m,n}\,,
~~
M_\te=-\sum_{mn\in\mZ }
s_{m,n}f\!\left[\!\begin{array}{c} m\\n\end{array}\!\right]\!\!(z)T_{m,n}\,,
\eq
where
\beq{4.7}
\vf\!\left[\!\begin{array}{c} m\\n\end{array}\!\right]\!\!(z)=
\bfe(n\te z)\phi((m+n\tau)\te,z)\,,
\eq
\beq{4.8}
f\!\left[\!\begin{array}{c} m\\n\end{array}\!\right]\!\!(z)=
\bfe(n\te z)\p_u\phi(u,z)|_{u=(m+n\tau)\te\,}\,,
\eq
and $\phi(u,z)$ is defined in (\ref{A.3}).
The equivalence of (\ref{4.5}) and (\ref{4.2}) follows from the Calogero
functional equation (\ref{ad2}).

Consider the holomorphic vector bundle $E$ of
infinite rank over $E_\tau$
with the structure group $G_\te$. We assume that it is similar to the
$\GLN$ bundle of degree one (see
Appendix A), where $\GLN$ is replaced by $G_\te$.
It means that the transition
functions $g_\al,\,\al=1,2$ have the form
$$
g_1=Q\,,~~g_2=\ti{\La}=\bfe((-\oh\tau+z)\te)\La\,.
$$
 The Higgs bundle is
$(T^*E,\, {\cal O}_{\cS^0})$, where the coadjoint orbit
${\cal O}_{\cS^0}$ is defined by (\ref{co}).
The cotangent bundle $T^*E$ is described by the Higgs field $\Phi=f^{-1}(z)L_\te f(z)$.
The Lax operator $L_\te$ satisfies the moment constraint equation
\beq{la}
\bp L_\te=0,~~ Res\,L_\te|_{z=0}=\cS\,,
\eq
\beq{qp}
L_\te(z+1)=QL_\te(z)Q^{-1}\,,
L_\te(z+\tau)=\La L_\te(z)\La^{-1}\,.
\eq
The reduced phase space is described by solutions of (\ref{la}), (\ref{qp})
such that
\beq{4.12}
I_{s,j}=\int_{E_\tau}\int_{{\cal A}_\te}(L_\te)^j\mu_{s,j}<\infty\,,
~~(s\leq j,\,j\in\mN)\,,
\eq
and $\mu_{s,j}$ are defined by (\ref{dif}) and (\ref{rep4}).
The integrals $ I^{er}_{s,j}$
can be extracted from the expansion over the basis of the
elliptic functions (\ref{rep2})
$$
\int_{{\cal A}_\te} (L_\te)^j(z)=I_{0,j}+
\sum_{r=2}^{j}I_{r,j}\wp^{(r-2)}(z),~(j=2,\ldots)\,.
$$
In particular,
$$
\int_{{\cal A}_\te} (L_\te)^2(z)=I_{0,2}+
\wp(z)\int_{{\cal A}_\te}\cS^2\,,~~I_{0,2}=2\pi^2\te^2H_\te\,.
$$
Note that
$$
I_{j,j}\sim C_j=\int_{{\cal A}_\te}\cS^j
$$
 are the Casimirs (\ref{Ca}).

Consider, for example, the integrals, that
 have the third order in the field $\cS$.
 It follows from (\ref{ad5})
that in terms of the Fourier modes $\cS=\{s_{m,n}\}$
the integrals take the form
\beq{4.16}
I_{2,3}=\sum_{\sum m_j=\sum n_j=0}
\prod_{j=1}^3s_{m_jn_j}
\left(
\ze\ar{m_1}{n_1}+\ze\ar{m_2}{n_2} +\ze\ar{m_3}{n_3}  \right)\,,
\eq
$$
(\ze\ar{m}{n}=\ze((m+n\tau)\te;\tau))\,,
$$
\beq{4.17}
I_{0,3}=
\sum_{\sum m_j=\sum n_j=0}
\prod_{j=1}^3s_{m_jn_j}\times
\eq
$$
\times\left(-\oh\wp'\ar{m_3}{n_3}-\wp\ar{m_3}{n_3}
[\ze\ar{m_1}{n_1}+\ze\ar{m_2}{n_2} +\ze\ar{m_3}{n_3}]
\right)  \,.
$$

The functionals (\ref{4.12}) play the role of Hamiltonians for
the infinite hierarchy on the phase space ${\cal O}_{\cS^0}$ (\ref{co})
\beq{4.13}
\p_{s,j}\cS=\{\nabla I_{s,j},\cS\}^*~~~(\p_{s,j}=\p_{t_{s,j}},
~~\nabla I^{er}_{s,j}=\frac{\de I^{er}_{s,j}}{ \de \cS})\,.
\eq
 It contains (\ref{4.4}), (\ref{4.2}) for $I_{0,2}$.

\bigskip
{\sl 3. Classical limit}.

In the classical limit $\cS$ becomes a function on $T^2$
$$
\cS=\sum_{m,n\in\mZ}s_{m,n}\exp(2\pi i(-mx-ny))\,.
$$
In our case  the classical limit essentially is the same as
the rational limit of the basic spectral curve $E_\tau$.
Replace for a moment the half periods $(\oh,\frac{\tau}{2})$  of $E_\tau$  on
 $\om_1,\om_2$.
The rational limit means that $\om_1,~\om_2\to\infty$.
The Weierstrass function degenerates as
$$
\wp(u)\to \f1{u^2}\,.
$$
Consider the double limit $\te\to 0$, $\om_1,~\om_2\to\infty$
 such that $\lim\om_1\te=1,~~\lim\om_2\te=\tau$.
Then
$$
\wp\!\left[\!\begin{array}{c} m\\n\end{array}\!\right]\to
\f1{(m+n\tau)^2}\,.
$$
 The quadratic Hamiltonian in the double limit takes the form
\beq{4.3a}
H=-\oh\int_{{\cal A}_\te}\psi (\bp)^2\psi=
-\oh\sum_{m,n\in\mZ}\frac{s_{m,n}s_{m,n}}{(m+n\tau)^2}\,,
\eq
where $\bp=\f1{2\pi i}(\p_x+\tau\p_y)$.
The operator $\bp^2$ plays the role of the inertia-tensor.
It replaces the Laplace operator $\Delta\sim\p\bp$
of the standard hydrodynamics. We call this system the {\sl modified
hydrodynamics}.

In the classical limit the Lax equation assumes the form
$$
\p_t L(x_1,x_2;z)=\{L(x_1,x_2;z),M(x_1,x_2)\}\,,
$$
where
\beq{8.10}
L(x,y;z)=\bp^{-1}\cS(x,y)+\f1{z}\cS(x,y)\,,
\eq
and
\beq{9.10}
M(x,y)=-\bp^{-2}\bfS(x,y)\,.
\eq

 The integrals of motion (\ref{4.12}) survive in this limit.
We already pointed the form of the Hamiltonian $H$ (\ref{4.3a}).
The third order integrals take the forms
$$
I_{2,3}=\sum_{\sum m_j=\sum n_j=0}
\prod_{j=1}^3s_{m_,jn_j}\sum_j\f1{m_j+n_j\tau}\,,
$$
$$
I_{0,3}=\sum_{\sum m_j=\sum n_j=0}
\prod_{j=1}^3s_{m_j,n_j}\times
$$
$$
\left(
-\f1{(m_3+n_3\tau)^3}-\f1{(m_3+n_3\tau)^2}[ \f1{m_1+n_1\tau}+
\f1{m_2+n_2\tau}+\f1{m_3+n_3\tau}]
\right) \,.
$$

\bigskip
{\sl 4. Reduction to the loop algebra}

Let $\te$ be a rational number $\te=p/N$. As it was explained in Section 2.3
we can pass
to the factor-algebra $L(\gln)$ or its central extension
$\hat{L}(\gln)$. In the first case
we have a family of non-interacting
ER parameterized by $w\in S^1$. If the central charge
nonzero  the situation is drastically changed \cite{LOZ,Kr}.
The Lax operator is no longer a one-form, but a connection on $S^1$
\beq{con}
\p_w+ L(z,w)\,,~~(w\in S^1)\,.
\eq
The integrals of motion can be calculated from the
expansion of the trace of the monodromy matrix for the linear system
$$
(\p_w+L(z,w))\Psi(z,w)=0\,.
$$
They define the hierarchy of the ER
on the coadjoint orbits of $\hat{L}(\GLN)$.
For $N=2$ this top is just the Landau-Lifshitz
equation
\beq{LL}
\p_t\cS=\frac{1}{2}[\cS,J(\cS)]+\oh[\cS,\p_{ww}\cS].
\eq
Here $\cS\in L^*(\slt)$.

\section{Elliptic Calogero-Moser system on ${\cal A}_\te$}
\setcounter{equation}{0}

{\sl 1. $\SLN$-Elliptic Calogero-Moser system (CM$_N$) }

The elliptic CM$_N$ system was first introduced in the quantum version
\cite{C} and then in the classical \cite{Mo}.
 We consider its generalization -
CM$_N$  system with spin. The elliptic CM$_N$ corresponds to
the trivial Higgs bundle over the elliptic curve $E_\tau$ \cite{GN}.
Its phase space is
\beq{b15}
{\mathcal R}^{CM_N} =\{\mC^{2N},\ti{\cal O}\}\,,
\eq
where $\ti{\cal O}={\cal O}//D$ is the symplectic quotient of the coadjoint
orbit of $\SLN$
\beq{orb}
{\cal O}=\{p\in\sln~|~p=h^{-1}p^0h,~h\in\SLN\,,p^0\in D\}
\eq
with respect to the action of the diagonal subgroup $D$ of $\SLN$.
The moment constraint imply that the diagonal matrix elements of the orbit
vanish $p_{jj}=0$.
The space ${\mathcal R}^{CM_N}$ has the same dimension
$\dim{\cal R}_{(0)}^{red}=N(N-1)$ as for the elliptic rotator.

The Poisson structure has the form
\beq{pb10}
\{v_j,u_k\}=\de_{j,k}\,,~~\{p_{k,l},p_{j,n}\}=\de_{j,l}p_{k,n}-\de_{n,k}p_{j,l}\,,
\eq
where
$=\overrightarrow{v}=(v_1,\ldots,v_N),~~\overrightarrow{u}=(u_1,\ldots,u_N)$
are canonical coordinates on $\mC^{2N}$.

The Hamiltonian, that has the second order with respect to the momenta
$\bfv$, has the form
 \beq{ha}
 H_2^{CM_N}=\oh\sum_{j=1}^Nv_j^2+
\sum_{j>k}p_{jk}p_{kj}\wp(u_j-u_k;\tau)\,.
 \eq
 It describes the interaction
of $N$ particles with complex coordinates $u_1,\ldots,u_N$  on the elliptic
curve $E_\tau$ (\ref{bsc}). The pair-wise potential is defined by the
Weierstrass function. The spin degrees of freedom $p_{jk}$ looks like
EAT with the inertia-tensor determined by $\wp(u_j-u_k;\tau)$,
but the corresponding phase subspace in contrast with standard
EAT is the symplectic quotient ${\cal O}//D$.

The equation of motion with respect to $H_2^{CM}$ has the Lax form
 $\p_t L^{CM_N}=[ L^{CM_N}, M^{CM_N}]$  with
\beq{8.2}
L^{CM_N}=P+X,~{\rm where}~P=\di(v_1,\ldots,v_N),
~~X_{jk}=p_{jk}\phi(u_j-u_k,z)\,,
\eq
($\phi$ is defined by (\ref{A.3})) and
\beq{8.3}
M^{CM_N}=-D+Y,~{\rm where}~D=\di(Z_1,\ldots,Z_N),~~Y_{jk}=y(u_j-u_k,z)\,,
\eq
$$
Z_j=\sum_{k\neq j}\wp(u_j-u_k) ,~~y(u,z)=\frac{\p \phi(u,z)}{\p u}\,.
$$
The equivalence of the Lax equation and the equations of motion is based
again on (\ref{ad2}) and (\ref{A.7a}).

We use relations from Appendix A to derive the elliptic CM$_N$ system
and its Lax representation via the Hitchin construction \cite{GN}.
If $d=$degree$(E^{st}_N)=0$, then the transition functions $g_\al$
can be gauge transformed  to the  constant matrices (\ref{B10}).
The Lax operator is a meromorphic matrix-valued one-form.
Its quasi-periodicity properties are defined by the transition
functions (\ref{B10})
$$
L^{CM_N}(z+1)= L^{CM_N}(z)\,,~~
L^{CM_N}(z+\tau)= \exp(\overrightarrow{u} )L^{CM_N}(z)\exp(-\overrightarrow{u})\,.
$$
It has a simple pole at $z=0$
such that
\beq{8.2b}
{\rm Res}_{z=0}(L^{CM_N}(z))=L_{-1}^{CM_N}= p\in\ti{\cal O}\,.
\eq

The integrals of motion $I_{s,j}$ (\ref{B12a}) produce CM$_N$ hierarchy
\beq{ch}
\p_{s,j} L^{CM_N}=[ L^{CM_N}, M^{CM_N}_{s,j}]\,.
\eq
The properties of  $M^{CM_N}_{s,j}$ can be  extracted from the
equations of motion (\ref{B13a})
$$
M^{CM_N}_{s,j}(z+1)=M^{CM_N}_{s,j}(z)\,,
$$
\beq{M}
M^{CM_N}_{0,j}(z)
-\exp(\overrightarrow{u}) M^{CM_N}_{0,j}(z+\tau)\exp(-\overrightarrow{u})
=2\pi i (L^{CM_N})^{j-1}-\p_{0,j}\overrightarrow{u}\,.
\eq
For $s\neq 0$ we have
$$
M^{CM_N}_{s,j}(z)
-\exp(\overrightarrow{u}) M^{CM_N}_{s,j}(z+\tau)\exp(-\overrightarrow{u})
=-\p_{s,j}\overrightarrow{u}\,,
$$
and the singular part of $M^{CM_N}_{s,j}(z)$ has the form
$$
(M^{CM_N}_{s,j}(z))_-=(L^{CM_N}(z))^{j-1}z^s)_-\,.
$$
In particular, $I_{0,2}= H_2^{CM}$ and $ M^{CM_N}_{0,2}=M^{CM_N}$ (\ref{8.3}).

Let $f(z)$ be the gauge transformation
that diagonalaized $g_2$. It is defined up to the
conjugation by a constant diagonal matrix. This remnant gauge freedom is
responsible for the symplectic reduction of the orbit $\tilde{\cal
O}=\mathcal{O}//D$.

\bigskip
{\sl 2. Equilibrium configuration}

We prove now that the following configuration of particles and spins
is an equilibrium set with respect to the Hamiltonian $ H_2^{CM_N}$ (\ref{ha}).
Consider $N=n^2$ particles and the orbit
variables enumerated by the pair of integer numbers $a,b=1,\ldots,n$
\beq{eq}
p_{a,b,\,c,d}=\nu\,,~~v_{a,b}=0\,,~~u_{a,b}=\frac{a+b\tau}{N}\,
~~a,b=\overline{1,n}\,.
\eq
From the identity
\beq{l1}
\wp(Nz|\tau)=\f1{N^2}\left[\wp(z|\tau)+ \sum_{j=1}^N(\sum_{k=1}^{N-1}
\wp(z+\frac{j+k\tau}{N}|\tau))+\wp(z+\frac{j}{N}|\tau)
\right] \,.
\eq
one obtains
\beq{l2}
\sum_{j=1}^N(\sum_{k=1}^{N-1}
\wp(\frac{j+k\tau}{N}|\tau))+\wp(\frac{j}{N}|\tau)=0\,.
\eq
It follows from (\ref{pv}) and (\ref{l1})  that
\beq{l3}
\wp'(\frac{j+k\tau}{N}|\tau)+
 \sum_{m\neq j,n\neq k}
\wp'(\frac{j+k\tau}{N}-\frac{m+n\tau}{N}|\tau)=0\,.
\eq
Then (\ref{l3}) implies that (\ref{eq}) is the equilibrium set
in ${\mathcal R}^{CM_N}$ with respect to $ H_2^{CM_N}$ (\ref{ha}).
Moreover, (\ref{l2}) means that
the Hamiltonian (\ref{ha}) vanishes at this point
\beq{eqh}
H_2^{CM}=0\,.
\eq
Note, that configuration (\ref{eq}) is preserved by the action of the higher
integrals $I_{s,k}$.

\bigskip
{\sl 3. Symplectic Hecke correspondence}

There exists a canonical transformation ({\em Symplectic Hecke correspondence})
 that defines the pass from CM$_N$ model related to the Higgs bundle of degree zero to ER
on $\GLN$ related to Higgs bundle of degree one \cite{LOZ}. It is a singular
gauge transformation $\Xi$ with a special form of its kernel. An eigen-vector
of the
residue $L_1^{CM_N}$ (\ref{8.2b}) is annihilated by the kernel. Then this
gauge transform
\beq{gt1}
L^{er}=\Xi^{-1}L^{CM}_{2D}\Xi\,.
\eq
preserves the order of the pole.
The matrix $\Xi$ has the following form.
Let $p^0=\di(p_1,\ldots,p_N)$ be the diagonal matrix defining the coadjoint
orbit (\ref{orb}) in the elliptic CM$_N$ system. Then $\Xi=\Xi(p_l)$ depends
on a choice of the eigen-value $p_l$.
Consider the following $(N\times N)$-
matrix $\tilde\Xi(z, u_1,\ldots,u_N;\tau)$~:
$$
\tilde\Xi_{ij}(z, \bfu;\tau) =
\theta{\left[\begin{array}{c}
\frac{i}N-\frac12\\
\frac{N}2
\end{array}
\right]}(z-Nu_j, N\tau )\,,
$$
where $\theta{\left[\begin{array}{c}
a\\
b\end{array}
\right]}(z, \tau )$ is the theta function with characteristics (\ref{A.30}).
Then
\beq{8.14}
\Xi(z, \bfu,p_l;\tau)=\tilde\Xi(z)\times{\rm diag}\left(\frac{(-1)^l}{p_l}
\prod_{j<k;j,k\ne l}
\vartheta(u_k-u_j,\tau)\right)\,.
\eq
Consider the case $N=2$. The phase space has dimension two,
since the orbit variables (\ref{orb}) are gauged away.
 Let $\nu^2$ be the value of the Casimir of the orbit.
 Then the transformation takes the form
\beq{S}
\left\{
\begin{array}{l}
S_1=-v\frac{\theta_{10}(0)}{\vartheta'(0)}
\frac{\theta_{10}(2u)}{\vartheta(2u)}-
\nu\frac{\theta_{10}^2(0)}{\theta_{00}(0)\theta_{01}(0)}
\frac{\theta_{00}(2u)\theta_{01}(2u)}{\vartheta^2(2u)}\,,
 \\
S_2=-v\frac{\theta_{00}(0)}{\sqrt{-1}\vartheta'(0)}
\frac{\theta_{00}(2u)}{\vartheta(2u)}-
\nu\frac{\theta_{00}^2(0)}{\sqrt{-1}\theta_{10}(0)\theta_{01}(0)}
\frac{\theta_{10}(2u)\theta_{01}(2u)}{\vartheta^2(2u)}\,,
\\
S_3=-v\frac{\theta_{01}(0)}{\vartheta'(0)}
\frac{\theta_{01}(2u)}{\vartheta(2u)}-
\nu\frac{\theta_{01}^2(0)}{\theta_{00}(0)\theta_{10}(0)}
\frac{\theta_{00}(2u)\theta_{10}(2u)}{\vartheta^2(2u)}\,,
 \\
\end{array}
\right.
\eq
where $\nu^2=\oh(S_1^2+S_2^2+S_3^2)$.

\bigskip
{\sl 4. Elliptic CM system on ${\cal A}_\te$}

Consider the limit $N\to\infty$ of CM$_N$ system CM$_\infty$
corresponding to ${\cal A}_\te$. We identify
 the coordinates of infinite number of particles in
$E_\tau$ with the diagonal matrix in GL$_\infty$
$$
\vec{u}=\di(\ldots ,u_{-N},\ldots,u_{-1},u_0,u_1,\ldots,u_N,\ldots)\,,
$$
and let
$$
\vec{v}=\di(\ldots ,v_{-N},\ldots,v_{-1},v_0,v_1,\ldots,v_N,\ldots)\,.
$$
be their momenta.

The Hamiltonian of CM$_\infty$ has the form
\beq{Ham}
H^{CM_\infty}=\oh (\vec{v},\vec{v})+
\sum_{j<k,j,k\in\mZ}p_{jk}p_{kj}\wp(u_j-u_k;\tau)\,,
\eq
where the orbit elements $p_{jk}$ is written
in the terms of the generators $E_{jk}$. In spite of the infinite number of
the particles on the torus $E_\tau$ the Hamiltonian $H^{CM_\infty}$
remains finite around
the equilibrium configuration (\ref{eq}), (\ref{eqh}).

The phase space ${\cal R}^{CM_\infty}$ of CM$_\infty$ has the similar form as
in the finite-dimensional case (\ref{b15})
$$
{\cal R}^{CM_\infty}=\{\mC^\infty\oplus\mC^\infty;\ti{\cal O}_\infty \}\,.
$$
Here
$\ti{\cal O}_\infty={\cal O}_\infty //D$
is the symplectic quotient of the coadjoint orbit with respect to the
Cartan subgroup $D\subset SIN_\te$ generated by $T_{m,0}$,
$m\in\mZ$.
The coadjoint orbit ${\cal O}_\infty\subset sin_\te^*$ of $SIN_\te\subset$
GL$_\infty$ is
$$
{\cal O}_\infty=\{p\in sin_\te^*~|~p=h^{-1}p^0h,~h\in SIN_\te\}\,.
$$
We assume that $\vec{v}\in\mC^\infty$ satisfies (\ref{v1}).
There are additional restrictions
coming from the finiteness of the integrals (\ref{int}) defined below.

In terms of coordinates on gl$_\infty^*$ the Poisson brackets on are
given by the similar formulae as (\ref{pb10})
\beq{pb1}
\{v_j,u_k\}=\de_{j,k}\,,
\eq
\beq{pb2}
\{p_{k,l},p_{j,n}\}=\de_{j,l}p_{k,n}-\de_{n,k}p_{j,l}\,.
\eq

We express the coordinates of the particles
in terms of the coordinates on $\cA_\te$\\
 $\overrightarrow{u}=\sum_{j\neq 0}\ti{u}_jT_{j,0}$
\beq{u}
u_j=\frac{i}{2\pi\te}\sum_{k\in\mZ}\ti{u}_k\bfe(jk\te)\,,~~\ti{u}_k\in\gS\,.
\eq
Evidently, $u_j$ is represented by the convergent series.
Similarly,
\beq{v}
v_j=-2\pi i\te^2\sum_{k\in\mZ}\ti{v}_k\bfe(jk\te)\,.
\eq
where $\ti{v}_k\in\gS'$ (\ref{ds}) and we assume that
\beq{v1}
\sum_{k\in\mZ}(v_k)^j<\infty\,,~~j=2,3\ldots\,.
\eq
Consider the generating functions
\beq{gf}
\bfu(x)=\frac{i}{2\pi\te}\sum_{m\in\mZ}\tilde{u}_m\bfe(x)^{m}\,,
\eq
and
\beq{gf1}
\bfv(x)=-2\pi i\te^2\sum_{m\in\mZ}\tilde{v}_m\bfe(x)^{-m}\,,
\eq
where we identify $\bfe(x)$ with the generator $U_1$.
In terms of coordinates $(\tilde{v}_m,\tilde{u}_l)$ the Poisson brackets takes the form
\beq{pb3}
\{\ti{v}_m,\ti{u}_n\}=\de_{m,n}\,,
\eq
or
\beq{pb4}
\{\bfv(x),\bfu(x')\}=\de(x-x')\,,
\eq
where $\de(x)=\sum_{m\in\mZ}\bfe(m\te x)$.

Define the orbit variables in the basis $T_{m,n}$:
\beq{pn}
{\cal S}(x,y)=-2\pi i \te^2\sum_{m,n}\bfe\left(\frac{mn\te}{2}\right)
s_{m,n}\bfe(x)^{-m}*\bfe(y)^{-n}\,,~~(U_1\sim\bfe(x)\,,~U_2\sim\bfe(y))\,.
\eq
It follows from (\ref{3.10}) that in terms of the coordinates on the NCT
$s_{m,n}$ the orbit variables are expanded as
\beq{p}
p_{j,j+n}=-2\pi i \te^2\sum_{m\in\mZ}\bfe\left(m\te(\frac{n}{2}-j)\right)s_{m,n}\,.
\eq
Since $p_{j,j}=0$, $s_{m,0}=0$ and $\cS(x,0)\equiv 0$.
The brackets (\ref{pb2}) takes the form
\beq{pb5}
\{s_{m,n},s_{m',n'}\}=\f1{\pi\te}\sin(\pi\te(mn'-m'n))s_{m+m',n+n'}\,.
\eq

The Hamiltonian (\ref{Ham}) can be rewritten
in terms of the NCT variables.
Using (\ref{u}) and (\ref{gf}) we find
$$
\wp(u_j-u_{j+n};\tau)=\wp(\bfu(\te j)-\bfu(\te (j+n));\tau)\,.
$$
Similarly to (\ref{p}) we define the coefficients $r_{m,n}$ as
$$
\wp(\bfu(\te j)-\bfu(\te (j+n)))=
\sum_{m\in\mZ}\bfe\left(m\te(\frac{n}{2}-j)\right)r_{m,n}\,.
$$
and the corresponding function on $\cA_\te$
$$
{\cal P}(x,y)=\sum_{m,n}\bfe\left(\frac{mn\te}{2}
\right)r_{m,n}\bfe(x)^{-m}*\bfe(y)^{-n}\,.
$$
Thus,
$$
\wp(u_j-u_{j+n};\tau)p_{j,j+n}=-2\pi \te^2\sum_{m}\bfe\left(m\te(\frac{n}{2}+j)
\right)
\sum_k s_{k-m,n}r_{k,n}\,.
$$
It allows us to put in the correspondence to  the  product $\wp(u_j-u_{j+n};\tau)p_{j,j+n}$
 the "convolution"
$$
({\cal P}\odot{\cal S})(x,y):=-2\pi \te^2\sum_{m,n}
\bfe(\te\frac{mn}{2})
\left(\sum_kr_{k,n}
s_{k-m,n}\right)\bfe(x)^{-m}*\bfe(y)^{-n}\,.
$$
Along with (\ref{gf1}) and (\ref{pn}) it leads to the
following expression for
the Hamiltonian (\ref{Ham})
$$
H^{CM_\infty}=\f1{2}\int_{{\cal A}_\te}\bfv(x)^2dx+\int_{{\cal A}_\te}
({\cal P}\odot{\cal S})(x,y)*\cS(x,y)\,.
$$

The CM$_\infty$ comes from the trivial infinite rank Higgs bundle  over
$E_\tau$ with transition functions $g(z)\in SIN_\te $. The whole procedure
is similar to finite-dimensional case. In particular,
$$
L^{CM_\infty}=P+X\,.
$$
Here
\beq{P}
P=\di(\ldots ,v_{-N},\ldots,v_{-1},v_0,v_1,\ldots,v_N,\ldots)\,,
\eq
\beq{b.17}
X_{jk}=p_{jk}\phi(u_j-u_k,z)\,~~p_{jk}\in\ti{\cal O}_\infty\,.
\eq
Define the coefficients $\ti{\phi}_{m,n}$ by the expansion
$$
\phi(u_j-u_{j+n};z)\equiv\phi(\bfu(\te j)-\bfu(\te
(j+n));z)=
\sum_{m}\bfe(m\te(\frac{n}{2}-j))\ti{\phi}_{m,n}(\bfu;z)\,.
$$
and construct the generating function
$$
{\cal F}(\bfu,x,y;z)=
\frac{i}{2\pi\te}\sum_{m,n}\ti{\phi}_{m,n}(\bfu;z)\bfe(\frac{mn\te}{2})\bfe(x)^{-m}
*\bfe(y)^{-n}\,.
$$
In the terms of the NCT $L^{CM_\infty}$ has the form
$$
L^{CM_\infty}(x,y)=\bfv(x)+({\cal S}\odot{\cal F}(\bfu,x,y;z))(x,y)\,,
$$
where $\bfv$ and $\cS$ are defined by (\ref{gf1}) and (\ref{pn}).

We have the infinite set of the integrals of motion
\beq{int}
I_{s,j}=\int_{E_\tau}\int_{\cA_\te} (L^{CM_\infty})^j\mu_{s,j}  \,,
\eq
and we assume that they are finite $ I_{s,j}<\infty$.
In particular,
$$
\int_{{\cal A}_\te} (L^{CM_\infty})^2(z)=
I_{0,2}+\wp(z)I_{2,2}\,,~~I_{2,2}=\int_{{\cal A}_\te}\cS^2\,,~~H^{CM_\infty}=\oh I_{0,2}\,.
$$

The integrals (\ref{4.12}) give rise to the hierarchy of the commuting
flows $\p_{s,j}\sim\{I_{s,j},~\}$.

\bigskip
{\sl 5. Reduction to the loop algebra.}

For a rational number $\te=p/N$ one can pass
to
$\hat{L}(\gln)$. The Lax operator being a one-form on $S^1$ (\ref{con})
acquires a form \cite{LOZ,Kr}
$$
L^{CM}=-\frac{\delta_{ij}}{2\pi\sqrt{-1}}\left(
\frac{v_i}{2}+\sum\limits_{\al}p^\al_{ii}E_1(z-w_\al)\right)-
\frac{1-\de_{ij}}{2\pi\sqrt{-1}}\sum\limits_{\al}p^\al_{ij}
\phi(u_{ij},z-w_\al)\,.
$$

The integrals of motion can be calculated from the
expansion of the trace of the monodromy matrix for the linear system
$$
(\p_w+L(z,w))\Psi(z,w)=0\,.
$$
They define the hierarchy of the elliptic CM field theory. For $N=2$ the first
non-trivial integral has the form
\beq{Ha}
H=\oint \frac{dw}{w}
\left(-\frac{v^2}{16\pi^2}(1-\frac{u_w^2}{h})
+(3u_w^2-h)\wp(2u)-\frac{u_{ww}^2}{4\nu^2}
\right)\,,
\eq
where $h$ is a Casimir corresponding to the co-adjoint orbit of $\hat{L}(\GLN)$ and
 $\nu^2=h-u_w^2$. For an arbitrary $N$ the quadratic Hamiltonians
of the type $I_{0,2}$ were calculated in Ref.\,\cite{Kr}.

 Let $L^{LL}$ be the Lax operator for the Landau-Lifshitz
equation and $L^{CM}_{2D}$ the Lax operator corresponding to
(\ref{Ha}). Then (see(\ref{gt1}))
$$
L^{LL}=\Xi^{-1}\p_w\Xi      +\Xi^{-1}L^{CM}_{2D}\Xi\,,
$$
where $\Xi$ is defined by (\ref{8.14}) for $N=2$. The explicit
relations between the phase space variables are given by (\ref{S}).

\section{Conclusion}
There are four related subjects that are not covered here.

$\bullet$
We have not considered here the classical limit of the CM$_\infty$
model. One can try to describe it independently as the Hitchin system with the
structure group SDiff$(T^2)$.

$\bullet$
It can be expected that the symplectic Hecke correspondence
survives in the limit $N\to\infty$. It would imply that CM
 system on the non-commutative torus ${\cal A}_\te$
and ER on ${\cal A}_\te$ are symplectomorphic.
It means in particular that the former system is not far from the non-commutative
modification of the 2D hydrodynamics. The symplectic Hecke correspondence
just boil the particles degrees of freedom to the orbit variables.
It can be suggested that the correspondence survives in the classical limit.

$\bullet$
It will be interesting to define the both systems on the central
extended algebra $\widehat{sin}_\te$ (\ref{3.13}). The central
charge produces the additional dimension and the corresponding
systems cover the CM field theory and the Landau-Lifshitz model.

$\bullet$
Two different tori are incorporated in our construction - the gauge
NCT ${\cal A}_\te$ and the basic spectral curve $E_\tau$.
In the classical limit they become dual. It seems natural to replace
 $E_\tau$ on another NCT  ${\cal A}_{\te'}$. In a general setting it means
a generalization on  the Higgs bundles over the non-commutative base.
The categories of holomorphic vector bundles on the non-commutative torus
were constructed in the recent paper \cite{SchP}.
 One attempt in this direction was done in Ref.\,\cite{T2}.

\section{Acknowledgments} The work was supported by the grant NSh-1999-2003.2 of the scientific schools
and RFBR-03-02-17554.

\section{Appendix}

\subsection{Appendix A. Hitchin systems on an elliptic curve.}
\setcounter{equation}{0}
\def\theequation{A.\arabic{equation}}

Let $E^{st}_N$ be a rank $N$ stable holomorphic vector bundle
over the elliptic curve $E_\tau$(\ref{bsc}).
It can be described by the holomorphic $\GLN$-valued transition functions
\beq{B1}
g_1(z):~z\to z+1\,,~~g_2(z):~z\to z+\tau\,,
\eq
$$
g_\al\in\Om^{0)}({\cal U_\al},{\rm Aut}^* E_N)~~\al=1,2\,,
$$
 $$
({\cal U}_1 {\rm ~is~ a~ neighborhood ~of~} [0,\tau]\,,~
{\cal U}_2 {\rm ~is~ a~ neighborhood ~of~} [0,1])\,.
$$
They satisfy the cocycle conditions
$$
g_1(z)g_2(z+1)g_1^{-1}(z+1+\tau)g_2^{-1}(z+\tau)=Id\,.
$$
Define the action of the gauge group ${\cal G}_N=\{f(z)\}$ as
\beq{B2}
g_1(z)\to f(z)g_1(z)f^{-1}(z+1)\,,~~
g_2(z)\to f(z)g_2(z)f^{-1}(z+\tau)\,.
\eq
The moduli space of the stable holomorphic bundles
 ${\cal M}_N(E_\tau)$ is defined as the quotient
\beq{B3}
{\cal M}_N(E_\tau)={\cal G}_N \backslash E_N\,.
\eq
The space ${\cal M}_N(E_\tau)$ is a disjoint union of the
components labeled by the corresponding degrees $d=c_1(\det E_N)$ :
\mbox{${\cal M}_N(E_\tau)=\bigsqcup{\cal M}_N^{(d)}$.}
The tangent space to  ${\cal M}(E_\tau)$ is isomorphic to
$h^1(E_\tau,{\rm End}E^{st}_N)$. Its dimension can be extracted from
 the Riemann-Roch theorem
$$
\dim h^1(E_\tau,{\rm End}E_N)= \dim h^0(E_\tau,{\rm End}E_N) \,.
$$
As a result we have
\beq{B4}
 \dim{\cal M}_N^{(d)}=g.c.d.(N,d)
\eq
The generic stable bundles can be transformed by (\ref{B2}) to the constant
diagonal form. For the trivial bundles $(d=0)$
\beq{B10}
g^{(0)}_1 =Id\,,~~g^{(0)}_2= f^{-1}(z) \di\exp\bfu f(z)\,.
\eq
For $d=1$ the transition functions can be chosen in the form
\beq{B15}
g^{(1)}_1= f^{-1}(z) Q_N f(z)\,,~~
g^{(1)}_2=  f^{-1}(z) \ti{\La}_N f(z) \,,
~~\ti{\La}_N =\bfe(-\frac{\oh\tau+z}{N})\La_N \,,
\eq
where
\beq{ql}
Q_N=\di(1,\bfe(\f1{N}),\ldots,\bfe(\f1{N-1}))\,,~~
\La_N=\sum_{j=1,N,~(mod~N)}E_{j,j+1}\,.
\eq

Consider the cotangent bundle $T^* E^{st}_N$.
We choose them in the following form.
$$
\eta_\al\in\Om^{(1,0)}({\cal U_\al},{\rm End}^* E^{st}_N)~~\al=1,2\,,
$$
 $$
({\cal U}_1 {\rm ~is~ a~ neiborhood ~of~} [0,\tau]\,,~
{\cal U}_2 {\rm ~is~ a~ neiborhood ~of~} [0,1])\,.
$$
The bundle $T^* E_N$ is called the {\sl Higgs bunle} over $E_\tau$.

We attribute to the marked point $z=0$ a coadjoint orbit of $\SLN$
\beq{B6}
{\cal O}=\{p\in\sln~|~p=h^{-1}p^0h,~h\in\SLN,~p^0\in\sln\}\,.
\eq
The unreduced phase space is the pair
\beq{b6}
{\cal R}=(T^* E^{st}_N ,{\cal O})
\eq
  with the symplectic form
\beq{B7}
\om=\sum_{\al=1,2}\oint_{\ga_\al}\tr(
D (g_\al^{-1}\eta_\al) \wedge Dg_\al)+\tr(D(h^{-1}p^0)\wedge Dh)\,.
\eq
Here the integrals $\oint$ is taken over
contours $\ga_1\sim [0,\tau),~\ga_2\sim [0,1)$. We assume that the marked point
$z=0$ lies inside the closed contour
$\ga_1(z)\ga_2(z+\tau)\ga_1^{-1}(z+1)\ga_2^{-1}(z)$.
The space ${\cal R}$ (\ref{b6}) is called the Higgs bundle
with the {\sl quasi-paraboloic structure} at the marked point $z=0$.

The canonical transformations of (\ref{B7}) are (\ref{B2}) along with
\beq{B8}
\eta_\al \to f(z)\eta_\al(z,\bz)f^{-1}(z),~~h\to hf(0)\,.
\eq
The transformations are generated by the following first class constraints.
Let $\Phi(z)$ be a meromorphic one-form on $E_\tau$. Then
\beq{B9}
\eta_\al=\Phi(z)\,,~~Res(\Phi(z))|_{z=0}=p\,,~ p\in{\cal O}\,.
\eq
The form $\Phi$ is the so-called {\sl the Higgs field}.
Moreover, the constraints imply the quasi-periodicity of $\eta_\al$
\beq{B5}
\eta_1(z,\bz)=g_1(z) \eta_1(z+1,\bz+1) g_1^{-1}(z)\,,~~
\eta_2(z,\bz)=g_2(z) \eta_2(z+\tau,\bz+\bar{\tau}) g_2^{-1}(z)\,.
\eq

Let $\mu_jd\bz\in\Om^{(-j,1)}(E_\tau)$ be $(-j,1)$-differentials on $E_\tau$.
We choose the representatives from $\Om^{(-j,1)}(E_\tau)$  that form
a basis in the cohomology space $h^1(E_\tau,\G^j)$, $(\dim h^1=j)$
\beq{dif}
\mu_j=(\mu_{0,j}\p_z^{j-1}, \mu_{2,j}\p_z^{j-1},
\ldots,\mu_{j,j}\p_z^{j-1})\,.
\eq
The coefficients $\mu_{s,j}$ coincide with the basis $f_s$ (\ref{rep4})
$(\mu_{s,j}= f_s )$.
The integrals
\beq{B12}
I_{s,j}=\int_{E_\tau}\tr(\Phi^j)\mu_{s,j}d\bz\,.
\eq
are gauge invariant.
The play the role of the Hamiltonians in the
integrable hierarchy. The equations of motion on the phase space ${\cal R}$
with respect to the Hamiltonian $ I_{s,j}$
take the form
\beq{B13}
\p_{s,j}\Phi=0\,,
\eq
\beq{B13a}
 (\p_{s,j}g_\al) g^{-1}_\al =\Phi^{j-1}\mu_{s,j}\,,
\eq
$$
\p_{s,j}p=0\,.
$$

Consider the symplectic quotient ${\cal R}^{red}={\cal R}//{\cal G}_N$. Let
$f(z)$ be the gauge transform that bring the transition function in the
standard form ((\ref{B10}) for $d=0$ and (\ref{B15}) for $d=1$).
The Lax operator is the corresponding gauge transform of the Higgs field $\Phi$
\beq{B11}
L_N(z)=f(z)\Phi(z)f^{-1}(z)\,.
\eq
Then the first equation (\ref{B13}) is equivalent to the Lax equation
\beq{lax}
\p_{s,j}L_N=[L_N,M_{N;s,j}]\,,
\eq
where $ M_{N;s,j}=f^{-1}\p_{s,j}f$.

In terms of the Lax matrix the integrals (\ref{B12}) have the form
\beq{B12a}
I_{s,j}=\oint\tr(L^{(0)}_N)^j \mu_{s,j}\,.
\eq
They can be found from the expansion on the basis of the elliptic funstions
\beq{B14a}
\tr(L_N)^j(z)=I_{0,j}+\sum_{s=2}^NI_{s,j}\wp^{(s-2)}(z)\,~~
(\wp^{(k)}(z)=\p_z^k\wp(z))\,.
\eq

\subsection{Appendix B. Elliptic functions.}
\setcounter{equation}{0}
\def\theequation{B.\arabic{equation}}

We summarize the main formulae for elliptic functions, borrowed
mainly from \cite{W} and \cite{Mu}.
We consider  the elliptic curve
\beq{bsc}
E_\tau=\mC/\mZ+\tau\mZ\,,~~
q=\bfe(\tau)=\exp 2\pi i\tau\,.
\eq

The basic element is the theta  function:
\beq{A.1a}
\vth(z|\tau)=q^{\frac
{1}{8}}\sum_{n\in {\bf Z}}(-1)^ne^{\pi i(n(n+1)\tau+2nz)}=
\eq
$$
q^{\frac{1}{8}}e^{-\frac{i\pi}{4}} (e^{i\pi z}-e^{-i\pi z})
\prod_{n=1}^\infty(1-q^n)(1-q^ne^{2i\pi z})(1-q^ne^{-2i\pi z})\,.
 $$
\bigskip

{\sl The Weierstrass functions}
\beq{A0}
\sigma(z|\tau)=\exp(\eta z^2)\frac{\vth(z|\tau)}{ \vth'(0|\tau)}\,,
\eq
where
\beq{A00}
\eta(\tau)=-\f1{6}\frac{\vth'''(0|\tau)}{ \vth'(0|\tau)}\,.
\eq
\beq{A.1}
\ze(z|\tau)=\p_z\log\vth(z|\tau)+2\eta(\tau)z\,, ~~
\ze(z|\tau) \sim\f1{z}+O(z^3)\,.
\eq

\beq{A.2}
\wp(z|\tau)=-\p_z\ze(z|\tau)\,.
\eq

\beq{4.0}
\wp(u;\tau)=\f1{u^2}+
+\sum'_{j,k}\left(
\f1{(j+k\tau+u)^2}-\f1{(j+k\tau)^2}\right)\,.
\eq

The next important function is
\beq{A.3}
\phi(u,z)=
\frac
{\vth(u+z)\vth'(0)}
{\vth(u)\vth(z)}\,.
\eq
It has a pole at $z=0$ and
\beq{A.3a}
\phi(u,z)=\frac{1}{z}+\zeta(u|\tau)+2\eta(\tau)u+
\frac{z}{2}((\zeta(u|\tau)+2\eta(\tau)u)^2-\wp(u))+\ldots\,,
\eq
and

{\sl Relation to the Weierstrass functions}
\beq{A3b}
\phi(u,z)^{-1}\p_u\phi(u,z)=\zeta(u+z)-\zeta(u)+2\eta(\tau)z\,.
\eq

\beq{A.7}
\phi(u,z)=\exp(-2\eta_1uz)
\frac
{\si(u+z)}{\si(u)\si(z)}\,,
\eq
\beq{A.7a}
\phi(u,z)\phi(-u,z)=\wp(z)-\wp(u)\,.
\eq

{\sl Particular values}
\beq{pv}
\wp'(z)=0~~{\rm for}~z=\oh,\frac{\tau}{2},\frac{1+\tau}{2}\,.
\eq

\bigskip

{\sl Addition formula. (Calogero functional equation.)}

\beq{ad2}
\phi(u,z)\p_v\phi(v,z)-\phi(v,z)\p_u\phi(u,z)=(\wp(v)-\wp(u))\phi(u+v,z)\,.
\eq

In fact, $\phi(u,z)$ satisfies more general relation which follows from the
Fay three-section formula
\beq{ad3}
\phi(u_1,z_1)\phi(u_2,z_2)-\phi(u_1+u_2,z_1)\phi(u_2,z_2-z_1)-
\phi(u_1+u_2,z_2)\phi(u_1,z_1-z_2)=0\,.
\eq
A particular case of this formula is (\ref{A.7a}) and
\beq{ad4}
\phi(u_1,z)\phi(u_2,z)-\phi(u_1+u_2,z)
(\zeta(u_1)+\zeta(u_2)-2\eta(\tau)(u_1+u_2))+
\p_z\phi(u_1+u_2,z)=0\,.
\eq
It follows from (\ref{A3b}), (\ref{A.7a}),(\ref{ad4}) that for $u_1+u_2+u_3=0$
\beq{4.15}
\phi(u_1,z)\phi(u_2,z)\phi(u_3,z)=
\left[\wp(z)-\wp(u_3)\right]
\left[\ze(u_1)+ \ze(u_2)+ \ze(u_3-z)+ \ze(z)\right]\,.
\eq
Then
\beq{ad5}
\phi(u_1,z)\phi(u_2,z)\phi(u_3,z)|_{z\to 0}=
 \f1{z^3}
+\f1{z^2}\left[\ze(u_1)+ \ze(u_2)+ \ze(u_3)\right]
\eq
$$
-\oh\wp'(u_3)-\wp(u_3)
\left[\ze(u_1)+ \ze(u_2)+ \ze(u_3)\right] +O(z)
 $$
\bigskip

  \bigskip
{\sl Basis of elliptic functions on $E_\tau$.}

We consider elliptic functions on $E_\tau$ with poles at $z=0$.
Any elliptic meromorphic function $F(z)$ is represented in the form
\beq{rep1}
F(z)=\sum_{j=0,2,3\ldots}c_je^j\,,
\eq
where
\beq{rep2}
e^0=1,~e^j=\p_z^{(j-2)}E_2(z)\,.
\eq
The dual basis $f_k$ with respect to the pairing
\beq{rep3}
\lan *|*\ran=\int_{E_\tau},~~\lan f_k|e^j\ran=\de^j_k
\eq
has the form
\beq{rep4}
f_0=(\bz-z)(1-\chi(z,\bz))\,,
\eq
$$
f_k=z^{k-1}\chi(z,\bz)\,,~k>1\,,
$$
where $\chi(z,\bz)$ is a characteristic function of a small neighborhood
${\cal U}_0 $ of $z=0$
\beq{cf}
\chi(z,\bz)=\left\{
\begin{array}{cl}
1,&\mbox{$z\in{\cal U}_0$ },~{\cal U}'_0\supset{\cal U}_0\\
0,&\mbox{$z\in E_\tau\setminus {\cal U}'_0$}\,.
\end{array}
\right.
\eq

\bigskip
{\sl Theta functions with characteristics.}\\
For $a, b \in \Bbb Q$ put~:
\beq{A.30}
\theta{\left[\begin{array}{c}
a\\
b
\end{array}
\right]}(z , \tau )
=\sum_{j\in \Bbb Z}
{\bf e}\left((j+a)^2\frac\tau2+(j+a)(z+b)\right)\,.
\eq
In particular, the  function $\vth$ (\ref{A.1a}) is the theta
function with a characteristic
\beq{A.29}
\vartheta(x,\tau)=\theta\left[
\begin{array}{c}
1/2\\
1/2
\end{array}\right](x,\tau)\,.
\eq

For the simplicity we denote $\theta\left[\begin{array}{l}
a/2\\b/2\end{array}\right]=\theta_{ab},$ $(a,b=0,1)$.

\end{document}